\documentclass[aps,preprint,showkeys,showpacs]{revtex4}%
\usepackage{amsfonts}
\usepackage{amsmath}
\usepackage{amssymb}
\usepackage{graphicx}%

\begin{document}
\title{Adiabatic Elimination in a Lambda System}
\author{E. Brion}
\email{ebrion@phys.au.dk}
\author{L.H. Pedersen}
\author{K. M\o lmer}
\affiliation{Lundbeck Foundation Theoretical Center for Quantum System Research, Department of Physics and Astronomy, University of Aarhus, 8000 Aarhus C, Denmark}

\begin{abstract}
This paper deals with different ways to extract the effective two-dimensional lower level dynamics of a lambda system excited by off-resonant laser beams. We present a commonly used procedure for elimination of the upper level, and we show that it may lead to ambiguous results. To overcome this problem and better understand the applicability conditions of this scheme, we review two rigorous methods which allow us both to derive an unambiguous effective two-level Hamiltonian of the system and to quantify the accuracy of the approximation achieved: the first one relies on the exact solution of the Schr\"{o}dinger equation, while the second one resorts to the Green's function formalism and the Feshbach projection operator technique.
\end{abstract}



\date{\today}

\maketitle

\section{Introduction}

When dealing with complicated multilevel systems, such as atoms or molecules,
it is necessary to look for allowed restrictions of the Hilbert space which
can lead\ to simplifications of the computational work. Thus, one usually
forgets states which are not populated initially and not coupled, either
\emph{directly} or \emph{indirectly}, to initially occupied states.

It is sometimes possible to go beyond this first step and isolate some subset
of initially occupied states if they are only \emph{weakly} and
\emph{non-resonantly} coupled to the others. The effective dynamics of such a
subset can then be approximately described by a Hamiltonian of smaller
dimensions than the original one, in which the effect of couplings outside the
relevant subspace is accounted for by additive energy shifts and couplings.
The procedure which allows one to get rid of the irrelevant states and derive
this effective Hamiltonian is called \emph{adiabatic elimination}. One of the
simplest examples of such a situation is the case of a three-level lambda
system, the low levels of which are initially populated and non-resonantly
coupled to the initially empty upper level via detuned harmonic perturbations: through adiabatic elimination, it is possible to reduce
the problem to an oscillating two-level system, which has been widely studied,
for instance in atomic physics \cite{AE87}.

The aim of the present paper is to understand how the adiabatic elimination
procedure works on this simple example and how it should be performed. After
briefly presenting the model (Sec. \ref{Model}), we show that an Ansatz
commonly used in the literature to adiabatically eliminate the excited state may lead to ambiguous results (Sec. \ref{RoughAdEl}). In order to better understand this scheme and explicit its conditions of applicability,
we then review two rigorous approximation methods to treat the problem: the first
one relies on the solution of the Schr\"{o}dinger equation and consists in neglecting the fast oscillating terms in the exact expression
of the amplitude of the excited state, which is then injected back into the
dynamical equations for the lower states (Sec. \ref{AdElRes}); the second one
resorts to the Green's function formalism and makes use of the pole approximation
which is discussed in detail (Sec. \ref{Green}).

\section{The lambda system\label{Model}}

\begin{figure}
[ptb]
\begin{center}
\includegraphics[
height=1.8 in,
width=3.9306in
]%
{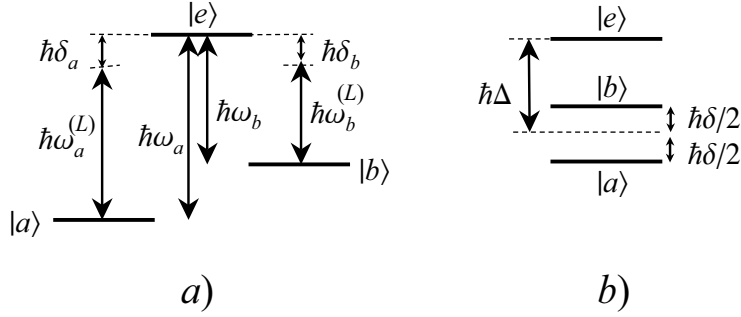}%
\caption{Lambda system's level scheme: a) in the Schrodinger picture, b) in the Rotating frame.}%
\label{Fig1}%
\end{center}
\end{figure}

In this paper, we shall consider an atomic lambda system\footnote{Note that, although we explicitly deal
with the case of a 3-level atom in laser fields, the model is fairly
general and can be applied to a wide range of physical systems.} consisting of two lower states $\left\{
\left\vert a\right\rangle ,\left\vert b\right\rangle \right\}  $ coupled to an excited level $\left\vert
e\right\rangle $ via two off-resonance lasers: the detunings are denoted $\delta_{k}=\omega
_{k}-\omega_{k}^{\left(  L\right)  }$ $\left(  k=a,b\right)  $ where
$\hbar\omega_{k}\equiv E_{e}-E_{k}$ and $\omega_{a,b}^{\left(  L\right)  }$
are the frequencies of the lasers (cf figure \ref{Fig1}a).

Turning to the rotating frame defined by the transformation $\left\vert
\psi\right\rangle \rightarrow\left\vert \varphi\right\rangle
=e^{+i\widehat{\xi}t}\left\vert \psi\right\rangle ,$ where
$
\widehat{\xi}\equiv\left[
\begin{array}
[c]{ccc}%
\frac{\delta}{2} & 0 & 0\\
0 & \omega_{a}-\omega_{b}-\frac{\delta}{2} & 0\\
0 & 0 & \omega_{a}-\frac{\delta_{a}+\delta_{b}}{2}%
\end{array}
\right]$ and $\delta\equiv\delta_{a}-\delta_{b} 
$, and performing the Rotating Wave
Approximation, one gets (cf figure \ref{Fig1}b)
\begin{equation}
\widehat{H}=\hbar\left[
\begin{array}
[c]{ccc}%
-\frac{\delta}{2} & 0 & \frac{\Omega_{a}^{\ast}}{2}\\
0 & \frac{\delta}{2} & \frac{\Omega_{b}^{\ast}}{2}\\
\frac{\Omega_{a}}{2} & \frac{\Omega_{b}}{2} & \Delta
\end{array}
\right]  , \label{Ham}%
\end{equation}
where $\Delta\equiv\frac{\delta_{a}+\delta_{b}}{2}$, and $\left( \Omega_{a}, \Omega_{b}\right)$ denote the Rabi frequencies of the lasers coupling
$\left\vert a\right\rangle $ to $\left\vert e\right\rangle $ and $\left\vert
b\right\rangle $ to $\left\vert e\right\rangle $, respectively. Note that we
have implicitly chosen the origin of the energies between the two states
$\left\vert a\right\rangle $ and $\left\vert b\right\rangle $ (see
figure \ref{Fig1}b).%

From now on, we shall assume $\left\vert \Delta\right\vert \gg\left\vert
\delta\right\vert ,\left\vert \Omega_{a}\right\vert ,\left\vert \Omega
_{b}\right\vert $. If the system is initially prepared in a superposition $\alpha_{0}\left\vert a\right\rangle +\beta_{0}\left\vert b\right\rangle $, the excited state will then essentially remain unpopulated, while second-order transitions will take place between the two lower states: this constitutes the so-called Raman transitions, which play an important role in atomic and molecular spectroscopy, and have recently become important processes in laser cooling and trapping \cite{DC89} and in quantum computing proposals with ions \cite{Wineland}, atoms \cite{Saffman}, and solid-state systems \cite{Ima}. In these conditions, it is natural to restrict the Hilbert space to the relevant states $\left\vert a\right\rangle $ and $\left\vert b\right\rangle $ and to describe their dynamics by a $2\times2$ effective Hamiltonian $H_{eff}$. In the following sections, we describe and discuss different ways to derive $H_{eff}$.

\section{Rough adiabatic elimination in a lambda system \label{RoughAdEl}}
The usual way to eliminate the excited state from the Schr\"{o}dinger equation, written for the state $\left\vert \psi\right\rangle =\left[
\begin{array}
[c]{l}%
\alpha\\
\beta\\
\gamma
\end{array}
\right]  $,
\begin{equation}
\left\{
\begin{array}
[c]{l}%
i\dot{\alpha}\left(  t\right)  =-\frac{\delta}{2}\alpha
+\frac{\Omega_{a}^{\ast}}{2}\gamma\\
i\dot{\beta}\left(  t\right)  =\frac{\delta}{2}\beta+\frac
{\Omega_{b}^{\ast}}{2}\gamma\\
i\dot{\gamma}\left(  t\right)  =\frac{\Omega_{a}}{2}\alpha
+\frac{\Omega_{b}}{2}\beta+\Delta\gamma
\end{array}
\right.  \label{ESnat}
\end{equation}
consists in claiming $\dot{\gamma}\left(  t\right)  =0$, which implies, by solving the last equation
\begin{equation}
\gamma=-\frac{\Omega_{a}}{2\Delta}\alpha-\frac{\Omega_{b}}{2\Delta
}\beta  \label{gamel}
\end{equation}
and injecting (\ref{gamel}) back into the dynamical equations for $\alpha$ and $\beta$, which yields
\begin{equation}
i\hbar\partial_{t}\left[
\begin{array}
[c]{l}%
\alpha\\
\beta%
\end{array}
\right]  =H_{eff}\left[
\begin{array}
[c]{l}%
\alpha\\
\beta%
\end{array}
\right] , 
\end{equation}
where the effective two-level Hamiltonian writes%
\begin{equation}
 H_{eff}=-\hbar\left[
\begin{array}
[c]{ll}%
\frac{\delta}{2}+\frac{\left\vert \Omega_{a}\right\vert ^{2}}{4\Delta} &
\frac{\Omega_{R}^{\ast}}{2}\\
\frac{\Omega_{R}}{2} & -\frac{\delta}{2}+\frac{\left\vert \Omega
_{b}\right\vert ^{2}}{4\Delta}%
\end{array}
\right], \quad \Omega_{R}\equiv \frac{\Omega_{a}\Omega_{b}^{\ast}}{2\Delta
}=\left\vert \Omega_{R}\right\vert e^{i\phi}.
\label{Hameff}
\end{equation}

If we had performed the same calculation in a shifted picture defined by the transformation
$\left\vert \varphi\right\rangle \rightarrow\left\vert \widetilde{\varphi}_{\eta
}\right\rangle =e^{-i\eta\Delta t}\left\vert \varphi\right\rangle \equiv \left[
\begin{array}
[c]{l}%
\widetilde{\alpha}_{\eta}\\
\widetilde{\beta}_{\eta}\\
\widetilde{\gamma}_{\eta}%
\end{array}
\right]  $, we would have obtained
\begin{equation}
\left\{
\begin{array}
[c]{l}%
i\dot{\widetilde{\alpha}}_{\eta}\left(  t\right)  =\left(
\eta\Delta-\frac{\delta}{2}\right)  \widetilde{\alpha}_{\eta}+\frac{\Omega
_{a}^{\ast}}{2}\widetilde{\gamma}_{\eta}\\
i\dot{\widetilde{\beta}}_{\eta}\left(  t\right)  =\left(
\eta\Delta+\frac{\delta}{2}\right)  \widetilde{\beta}_{\eta}+\frac{\Omega
_{b}^{\ast}}{2}\widetilde{\gamma}_{\eta}\\
i\dot{\widetilde{\gamma}}_{\eta}\left(  t\right)  =\frac{\Omega
_{a}}{2}\widetilde{\alpha}_{\eta}+\frac{\Omega_{b}}{2}\widetilde{\beta}_{\eta
}+\left(  1+\eta\right)  \Delta\widetilde{\gamma}_{\eta}.
\end{array}
\right.  
\end{equation}
and, applying the same Ansatz $\dot{\widetilde{\gamma}}_{\eta}\left(
t\right)  =0$ as before (we assume $\eta\neq-1$), we would have derived
\begin{equation}
\widetilde{\gamma}_{\eta}=-\frac{\Omega_{a}}{2\Delta\left(  1+\eta\right)
}\widetilde{\alpha}_{\eta}-\frac{\Omega_{b}}{2\Delta\left(  1+\eta\right)
}\widetilde{\beta}_{\eta} \label{gamshift}%
\end{equation}
which yields the effective Hamiltonian $\widetilde{H}_{eff,\eta}$ in the shifted picture. Finally, subtracting $\eta \Delta$ from $\widetilde{H}_{eff,\eta}$ we would have obtained the effective Hamiltonian $H_{eff,\eta}$ 
\[
 H_{eff,\eta}=-\hbar\left[
\begin{array}
[c]{cc}%
\frac{\delta}{2}+\frac{\left\vert \Omega_{a}\right\vert ^{2}}{4\Delta\left(
1+\eta\right)  } & \frac{\Omega_{R,\eta}^{\ast}}{2}\\
\frac{\Omega_{R,\eta}}{2} & -\frac{\delta}{2}+\frac{\left\vert \Omega
_{b}\right\vert ^{2}}{4\Delta\left(  1+\eta\right)  }%
\end{array}
\right], \quad \Omega_{R,\eta}\equiv\frac{\Omega_{a}\Omega_{b}^{\ast}}
{2\Delta\left(  1+\eta\right)  }=\left\vert \Omega_{R,\eta}\right\vert
e^{i\phi_{\eta}}
\]

We are thus led to the obviously unphysical conclusion that the effective Hamiltonian, $i.e.$ the effective dynamics of the system depends on the picture where the elimination is performed. This raises questions about the Ansatz we used: in which picture, if any, does it apply, and what is its physical meaning ? To understand better when and how to employ this scheme, we investigate two rigorous elimination methods in the next sections, which yield both an unambiguous expression of the effective Hamiltonian and the level of accuracy of the approximation achieved.

\section{Rigorous adiabatic elimination in a lambda system through solution of the Schr\"{o}dinger equation for the amplitude of the excited
state \label{AdElRes}}

In this section, we propose a straightforward elimination scheme based on the analysis of the exact expression of the amplitude $\gamma$ of the excited state: resorting to a simple mathematical argument, we identify its relevant part $\gamma_{rel}$ which mainly contributes to the dynamics of $\left( \alpha, \beta \right)$; injecting it back into the dynamical equations, we then derive an unambiguous expression for $H_{eff}$ which yields an approximation of the dynamics of the system, the accuracy of which can be quantified. This method moreover allows us to specify the applicability conditions of the previous scheme.

\subsection{Exact solution of the problem \label{ExactRes}}

\begin{figure}
[ptb]
\begin{center}
\includegraphics[
height=2.9 in,
width=3.2in
]%
{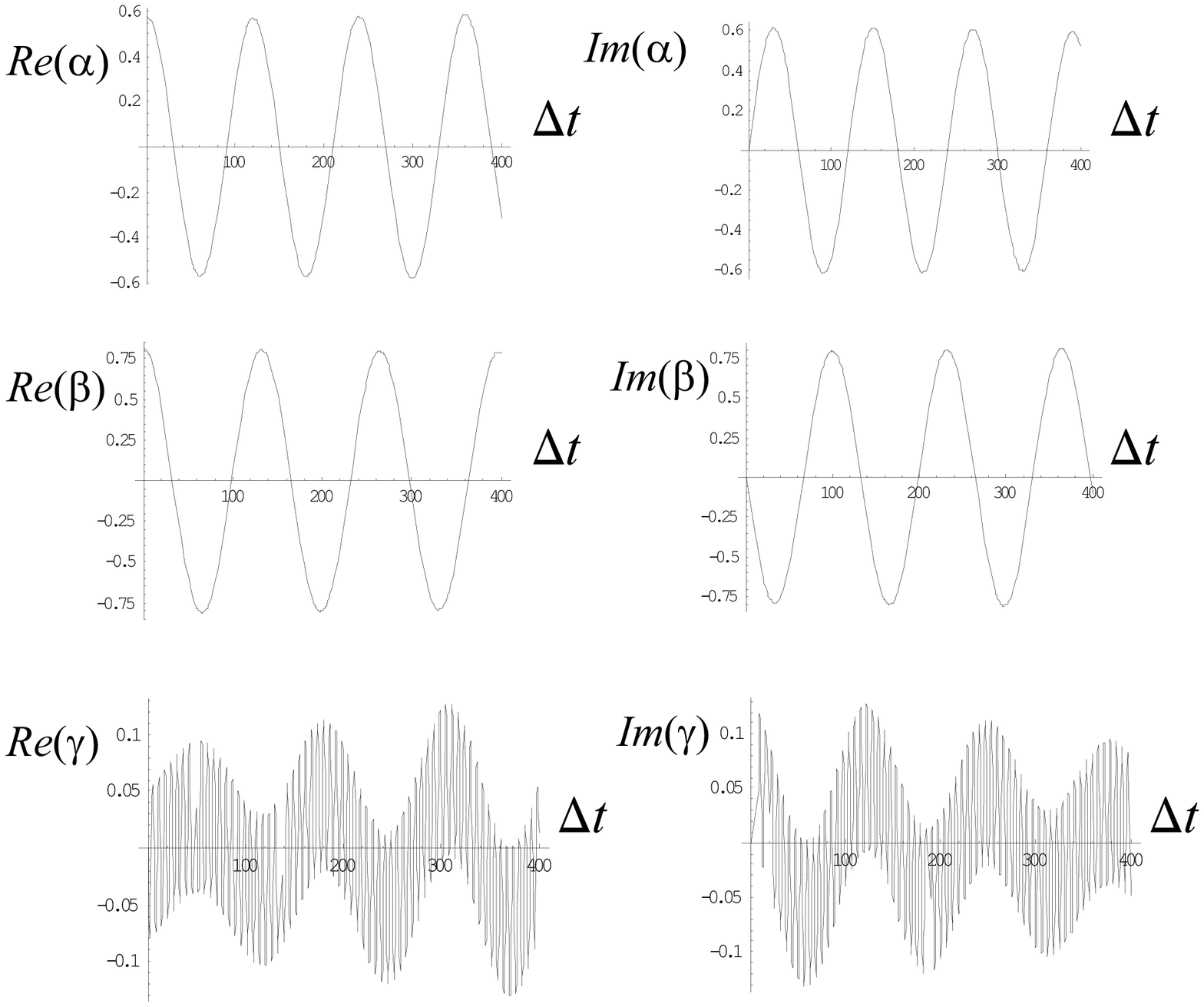}
\caption{Exact solutions for $\alpha$, $\beta$\ and $\gamma$.\ The values of
the parameters used for the simulation are: $\alpha_{0}=\sqrt{1/3}$, $\beta_{0}=\sqrt{2/3}$,
$\delta / \Delta=0.1$, $\Omega_{1}/\Delta =0.1 \times e^{-i\pi/3}$, $\Omega_{2}/\Delta = 0.1 \times e^{-i\pi/2}$.}
\label{Fig4}%
\end{center}
\end{figure}

\begin{figure}
[ptb]
\begin{center}
\includegraphics[
height=3in,
width=5in
]%
{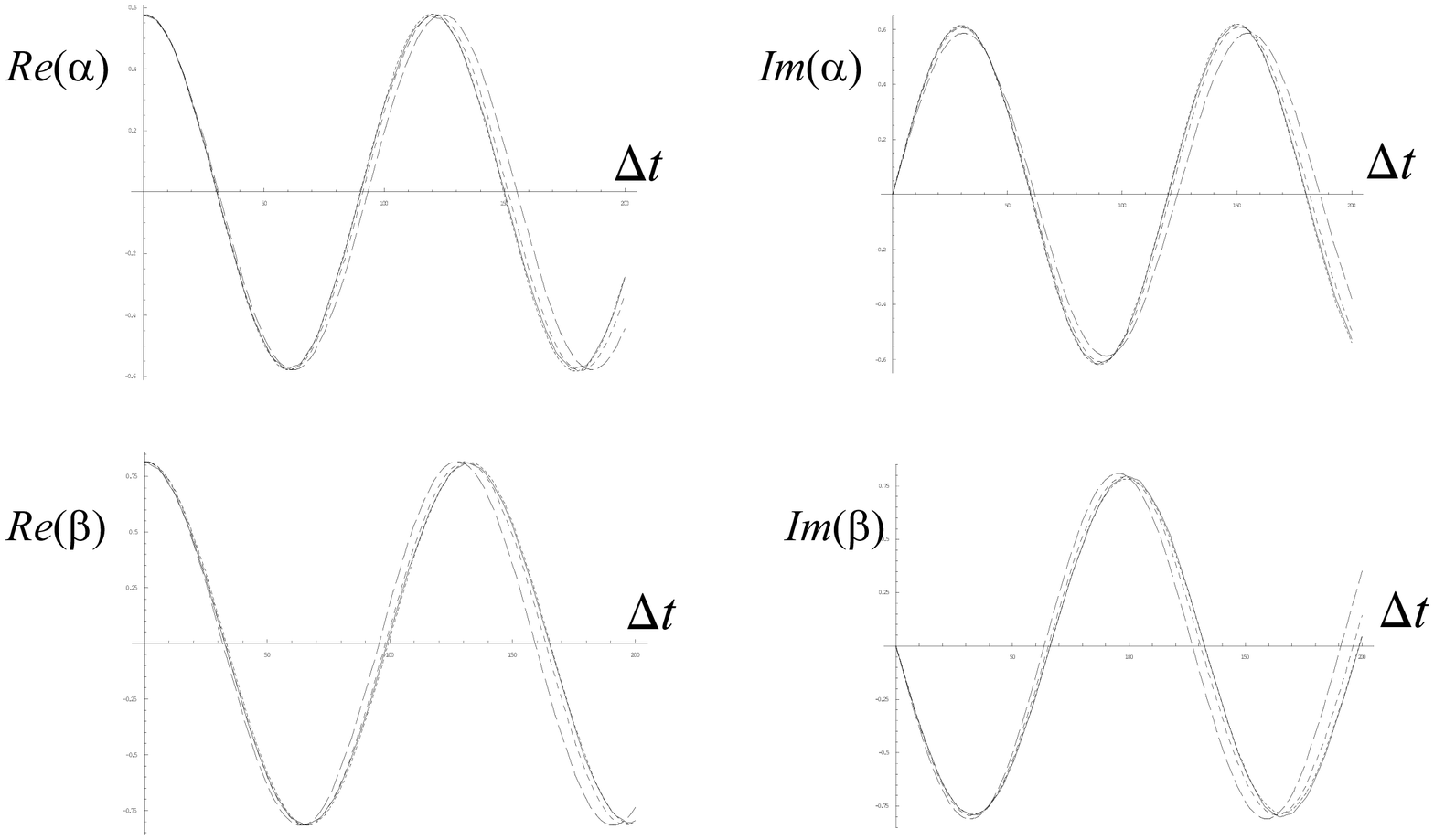}%
\caption{Comparison of the exact solutions with the rough approximation
calculated in a shifted picture, for different values of the shift $\eta$. The values of the parameters used for the simulation are:
$\alpha_{0}=\sqrt{1/3}$, $\beta_{0}=\sqrt{2/3}$,
$\delta / \Delta=0.1$, $\Omega_{1}/\Delta =0.1 \times e^{-i\pi/3}$, $\Omega_{2}/\Delta = 0.1 \times e^{-i\pi/2}$; exact solution (full line), $\eta=0$ (dashed line), $\eta=0.3$ (broken line), $\eta=3$  (longbroken line).\ }
\label{Fig5}%
\end{center}
\end{figure}

The derivation of the exact solutions of the Schr\"{o}dinger equation can be
straightforwardly performed by finding the eigenenergies of the system and
using the boundary condition $\left\vert \psi\left(  t=0\right)  \right\rangle
=\alpha_{0}\left\vert a\right\rangle +\beta_{0}\left\vert b\right\rangle $ to
determine the coefficients of the Fourier decompositions of the different
amplitudes. We do not reproduce these calculations but only
summarize the results which are useful for our purpose. Introducing the reduced variables $\left( \lambda,\lambda_{k=a,b} \right)$ such that $\lambda\epsilon\equiv\frac{\delta}{2\Delta}$, $\lambda_{k=a,b}\epsilon\equiv\frac{\Omega_{k}}{2\Delta}$,
with $0<\epsilon\ll1$, $\left( \lambda,\lambda_{k=a,b} \right) =O\left(  1\right)  $, and
$\lambda\geq0$\footnote{Note that $\left(\lambda,\lambda_{k=a,b},\epsilon \right)$ are not uniquely defined by these relations: the point is only to identify a common infinitesimal parameter $\epsilon$ for systematic expansion.}, one readily shows
\begin{equation}
\left\{
\begin{array}
[c]{l}%
\alpha\left(  t\right)  =\sum_{k=1}^{3}A_{k}e^{-i\Delta x_{k}t}\\
\beta\left(  t\right)  =\sum_{k=1}^{3}B_{k}e^{-i\Delta x_{k}t}\\
\gamma\left(  t\right)  =\sum_{k=1}^{3}C_{k}e^{-i\Delta x_{k}t}%
\end{array}
\right.  \label{ExactSol}%
\end{equation}
where $\left\{  x_{k}\right\}_{k=1,2,3}  $ are the solutions of the equation
\begin{equation}
 x^{3}-x^{2}-\left(  \lambda^{2}+\left\vert \lambda_{a}\right\vert
^{2}+\left\vert \lambda_{b}\right\vert ^{2}\right)  \epsilon^{2}x+\lambda
^{2}\epsilon^{2}+\lambda\epsilon^{3}\left(  \left\vert \lambda_{a}\right\vert
^{2}-\left\vert \lambda_{b}\right\vert ^{2}\right)  =0 \label{Eqfreq}%
\end{equation}
and the coefficients $\left\{ A_{k},B_{k},C_{k} \right\}_{k=1,2,3}$ are determined by the boundary conditions. Table \ref{TabRes} displays the expansions in $\epsilon$ of these different parameters.
\begin{table}
\caption{\label{TabRes}Expansions of the different parameters of the exact solutions.}
\begin{center}
\item[]
\begin{tabular}{|c||c|c|}
\hline
 & $\lambda = 0$ & $\lambda \neq 0$ \\ 
\hline 
$x_{1}$ & $O\left(  \epsilon^{2}\right)$ & $-\lambda\epsilon + O\left(  \epsilon^{2}\right)$ \\ 
\hline 
$x_{2}$ & $0$ & $\lambda\epsilon + O\left(\epsilon^{2}\right)$ \\ 
\hline 
$x_{3}$ & $1+O\left(  \epsilon^{2}\right)$  & $1+ O\left(  \epsilon^{2}\right)$ \\ 
\hline 
$A_{1}$ & $ \frac{\alpha_{0} \left\vert \lambda_{a} \right\vert^{2} + \beta_{0} \lambda_{a}^{\ast} \lambda_{b}}{\left\vert \lambda_{a} \right\vert^{2}+\left\vert \lambda_{b} \right\vert^{2}} + O\left( \epsilon^{2} \right)$ & $\alpha_{0} + \frac{\beta_{0} \lambda_{a}^{\ast} \lambda_{b}}{2 \lambda}\epsilon + O\left( \epsilon^{2} \right)$ \\ 
\hline 
$A_{2}$ & $\frac{\alpha_{0}\left\vert \lambda_{b}\right\vert ^{2}-\lambda
_{a}^{\ast}\lambda_{b}\beta_{0}}{\left\vert \lambda_{a}\right\vert
^{2}+\left\vert \lambda_{b}\right\vert ^{2}}$ & $ - \frac{\beta_{0} \lambda_{a}^{\ast} \lambda_{b}}{2 \lambda}\epsilon + O\left( \epsilon^{2} \right)$ \\ 
\hline 
$A_{3}$ & $O\left( \epsilon^{2} \right)$  & $O\left( \epsilon^{2} \right)$  \\ 
\hline 
$B_{1}$ & $ \frac{\alpha_{0} \lambda_{a} \lambda_{b}^{\ast}  + \beta_{0} \left\vert \lambda_{b} \right\vert^{2} }{\left\vert \lambda_{a} \right\vert^{2}+\left\vert \lambda_{b} \right\vert^{2}} + O\left( \epsilon^{2} \right)$ & $ \frac{\alpha_{0} \lambda_{a} \lambda_{b}^{\ast}}{2 \lambda} \epsilon + O\left( \epsilon^{2} \right)$ \\ 
\hline 
$B_{2}$ & $\frac{-\lambda_{a}\lambda_{b}^{\ast}\alpha_{0}+\left\vert
\lambda_{a}\right\vert ^{2}\beta_{0}}{\left\vert \lambda_{a}\right\vert
^{2}+\left\vert \lambda_{b}\right\vert ^{2}}$ & $\beta_{0} - \frac{\alpha_{0} \lambda_{a} \lambda_{b}^{\ast}}{2 \lambda} \epsilon + O\left( \epsilon^{2} \right)$ \\ 
\hline 
$B_{3}$ & $O\left( \epsilon^{2} \right)$  & $O\left( \epsilon^{2} \right)$  \\ 
\hline 
$C_{1}$ & $-\left(  \lambda_{a}\alpha_{0}+\lambda_{b}\beta_{0}\right)
\epsilon+O\left(  \epsilon^{2}\right)$ & $-\lambda_{a}\alpha_{0}\epsilon+O\left(  \epsilon^{2}\right)$ \\ 
\hline 
$C_{2}$ & $0$ & $-\lambda_{b}\beta_{0}\epsilon+O\left(  \epsilon^{2}\right)$ \\ 
\hline 
$C_{3}$ & $ \left(  \lambda_{a}\alpha_{0}+\lambda_{b}\beta_{0}\right)  \epsilon+O\left(
\epsilon^{2}\right)$ & $\left(  \lambda_{a}\alpha_{0}+\lambda_{b}
\beta_{0}\right)  \epsilon+O\left(  \epsilon^{2}\right)$ \\  
\hline
\end{tabular}
\end{center}
\end{table}
The rigorous solutions are represented in figure \ref{Fig4}, in the
regime $\left\vert \Delta\right\vert \gg\left\vert \delta\right\vert
,\left\vert \Omega_{a}\right\vert ,\left\vert \Omega_{b}\right\vert $: the
amplitudes $\alpha$ and $\beta$ show an oscillating behavior, which, as
expected, is much alike a two-level Rabi oscillation; $\gamma$ oscillates much
faster than $\alpha$ and $\beta$.
The comparison between the exact solutions and the approximations
obtained in the previous subsection (see figure \ref{Fig5}) shows that
the previous elimination method provides a valuable approximation only in what we called "the
natural picture". In the following, we clarify why this is so through a straightforward and rigorous elimination scheme.

\subsection{Straightforward adiabatic elimination \label{RigAdEl}}

\subsubsection{Preliminary remark \label{Prel}}

Consider the differential equation%
\begin{equation}
i\dot{f}\left(  t\right)  \pm \frac{\delta}{2}f\left(  t\right)
=\frac{\Omega}{2}\left(  A_{\omega}e^{-i\omega t}+A_{\Delta}e^{-i\Delta
t}\right) \label{eqnat}
\end{equation}
the solution of which writes
\[ 
f\left(  t\right)  =f\left(  0\right)  e^{\pm i\frac{\delta}{2}t}+\frac{\Omega
A_{\omega}}{2\omega \pm \delta}\left(  e^{-i\omega t}-e^{\pm i\frac{\delta}{2}%
t}\right)  +\frac{\Omega A_{\Delta}}{2\Delta \pm \delta}\left(  e^{-i\Delta
t}-e^{\pm i\frac{\delta}{2}t}\right)  .
\]
If $\omega=\lambda_{\omega}\Delta\epsilon$, $\Omega=2\lambda_{\Omega}\Delta\epsilon$ and $\delta=2\lambda\Delta\epsilon$, 
with $\left\vert \epsilon\right\vert \ll1$, $\left( \lambda_{\omega},\lambda_{\Omega
},\lambda \right) = O\left(1\right)$ and $\left(A_{\omega},A_{\Delta}\right)=O\left(  \epsilon\right)  $, one gets%
\[
 f\left(  t\right) \simeq  \stackrel{O\left(  1\right)  }{\overbrace{f\left(  0\right)e^{\pm i\frac{\delta}{2}t}}}  + \stackrel{O\left(  \epsilon\right)  }{\overbrace{\frac{\lambda_{\Omega}A_{\omega}}{\lambda_{\omega} \pm \lambda}\left(
e^{-i\omega t}-e^{\pm i\frac{\delta}{2}t}\right)  }}+\stackrel{O\left(
\epsilon^{2}\right)  }{\overbrace{\lambda_{\Omega}A_{\Delta}\epsilon\left(
e^{-i\Delta t}-e^{\pm i\frac{\delta}{2}t}\right)  }}+O\left(  \epsilon
^{3}\right).
\]
If one is interested in a solution valid to first order in $\epsilon$, the $O\left( \epsilon^{2} \right)$ term can be neglected, or, equivalently, one can forget the
corresponding term directly in the differential equation which then reduces to
\
\[
i\dot{f}\left(  t\right)  \pm \frac{\delta}{2}f\left(  t\right)
=\frac{\Omega}{2}A_{\omega}e^{-i\omega t}.
\]

In the same way, one straightforwardly shows that the solutions of the equation
\begin{equation}
i\dot{f}\left(  t\right)  + \left(\pm \frac{\delta}{2} - \eta \Delta  \right) f\left(  t\right)
=\frac{\Omega}{2}\left(  A_{\omega}e^{-i \left( \omega + \eta \Delta \right) t }+A_{\Delta}e^{-i\Delta \left(1 + \eta \right)
t}\right) \label{eqshift}
\end{equation}
where $\eta$ is an arbitrary real number and all the other parameters obey the same relations as before, coincide (to order $O\left( \epsilon \right)$) with those of the simpler equation
\[
i\dot{f}\left(  t\right)  + \left( \pm \frac{\delta}{2} - \eta \Delta  \right) f\left(  t\right)
=\frac{\Omega}{2}  A_{\omega}e^{-i \left( \omega + \eta \Delta \right) t } .
\]

\subsubsection{Rigorous adiabatic elimination of the excited state}
Let us now return to our initial problem. Injecting the exact expression (\ref{ExactSol}) of $\gamma$ into (\ref{ESnat}), we get the equations
\begin{eqnarray*}
i\dot{\alpha}  &  = & -\frac{\delta}{2}\alpha+C_{1}\frac{\Omega_{a}%
}{2}e^{-i\stackrel{\simeq-\frac{\delta}{2}\ll\Delta}{\overbrace{x_{1}\Delta}}%
t}+C_{2}\frac{\Omega_{a}}{2}e^{-i\stackrel{\simeq\frac{\delta}{2}\ll\Delta
}{\overbrace{x_{2}\Delta}}t}+C_{3}\frac{\Omega_{a}}{2}e^{-i\stackrel
{\simeq\Delta}{\overbrace{x_{3}\Delta}}t}\\
i\dot{\beta}  &  = & \frac{\delta}{2}\beta+C_{1}\frac{\Omega_{b}}%
{2}e^{-ix_{1}\Delta t}+C_{2}\frac{\Omega_{b}}{2}e^{-ix_{2}\Delta t}+C_{3}%
\frac{\Omega_{b}}{2}e^{-ix_{3}\Delta t}.
\end{eqnarray*}
which are of the form (\ref{eqnat}), with 
\[
\lambda_{\omega}=\pm\frac{\delta}{2\Delta}=\pm\lambda, \qquad%
\lambda_{\Omega}=\lambda_{a,b}.
\]
Neglecting the last term of each of these equations, or equivalently, replacing $\gamma$ by its
"relevant" component\footnote{$\gamma_{rel}$ can easily be shown to be equal, up to second order terms in $\epsilon$,
to the average of $\gamma$ over the period of its highest harmonic component
$\omega_{3}\simeq\Delta$.}
\[
\gamma_{rel}\left(  t\right)  \equiv\gamma\left(  t\right)  -C_{3}%
e^{-ix_{3}\Delta t}=\sum_{k=1}^{2}C_{k}e^{-ix_{k}\Delta t}
\]
then leads to approximate forms for $\left( \alpha, \beta \right)$, valid to order $O\left( \epsilon \right)$ terms. Let us now relate $\gamma_{rel}$ to $\alpha,\beta$: we have 
\begin{eqnarray*}
i\dot{\gamma}\left(  t\right)   &  = & i\dot{\gamma}%
_{rel}\left(  t\right)  +x_{3}\Delta C_{3}e^{-ix_{3}\Delta t}\\
& = & i\dot{\gamma}_{rel}\left(  t\right)  -x_{3}\Delta\left[
\gamma\left(  t\right)  -C_{3}e^{-ix_{3}\Delta t}\right]  +x_{3}\Delta
\gamma\left(  t\right) \\
& = & i\dot{\gamma}_{rel}\left(  t\right)  -x_{3}\Delta\gamma
_{rel}+x_{3}\Delta\gamma\left(  t\right) \\
& = & \frac{\Omega_{a}}{2}\alpha+\frac{\Omega_{b}}{2}\beta+\Delta\gamma
\end{eqnarray*}
whence
\begin{equation}
i\frac{\dot{\gamma}_{rel}\left(  t\right)  }{\Delta}+\left(
x_{3}-1\right)  \left(  \gamma-\gamma_{rel}\right) = \gamma_{rel}+\left(
\lambda_{a}\alpha+\lambda_{b}\beta\right)  \epsilon\label{gamrel}%
\end{equation}
and, as
\begin{eqnarray*}
i\frac{\dot{\gamma}_{rel}\left(  t\right)  }{\Delta}  &
= & \sum_{k=1}^{2}\stackrel{O\left(  \epsilon^{2}\right)  }{\overbrace{x_{k}}%
}\stackrel{O\left(  \epsilon\right)  }{\overbrace{C_{k}}}e^{-ix_{k}\Delta
t}=O\left(  \epsilon^{3}\right) \\
x_{3}-1  & = & \left(  \left\vert \lambda_{1}\right\vert ^{2}+\left\vert
\lambda_{2}\right\vert ^{2}\right)  \epsilon^{2}+O\left(  \epsilon^{3}\right)
\\
\gamma_{rel},\gamma & = & O\left(  \epsilon\right) \\
\alpha,\beta & = & O\left(  1\right)
\end{eqnarray*}
we deduce from (\ref{gamrel}) that
\begin{equation}
\gamma_{rel}=-\frac{\Omega_{a}}{2\Delta}\alpha-\frac{\Omega_{b}}{2\Delta}%
\beta + O\left( \epsilon^{2} \right). \label{GamRelRig}%
\end{equation}
This expression coincides with (\ref{gamel}): it means that, in the natural picture, the rough adiabatic elimination procedure indeed leads to a valid approximation of the actual dynamics of the system, to order $O\left( \epsilon \right)$, which agrees with the remark we made about figure \ref{Fig5} and legitimates the expression (\ref{Hameff}) for the effective Hamiltonian.

If we turn to the shifted picture defined by the transformation $\left\vert \varphi\right\rangle \rightarrow\left\vert \widetilde{\varphi}_{\eta
}\right\rangle =e^{-i\eta\Delta t}\left\vert \varphi\right\rangle $, the exact expression of the amplitude of the excited state is now
\[
\widetilde{\gamma}\left( t \right)= \sum_{k=1}^{3} C_{k} e^{-i \Delta \left( x_{k} + \eta \right) t }
\]
and leads to the dynamical equations
\begin{eqnarray*}
 i \dot{\widetilde{\alpha }} &=&\left( \eta \Delta -\frac{\delta }{%
2}\right) \widetilde{\alpha }+C_{1}\frac{\Omega _{a}}{2}e^{-i\left(
x_{1}+\eta \right) \Delta t}+C_{2}\frac{\Omega _{a}}{2}e^{-i\left(
x_{2}+\eta \right) \Delta t}+C_{3}\frac{\Omega _{a}}{2}e^{-i\left(
x_{3}+\eta \right) \Delta t} \\
 i \dot{\widetilde{\beta }} &=&\left( \eta \Delta +\frac{\delta }{2}\right) \widetilde{\beta }+C_{1}\frac{\Omega _{b}}{2}e^{-i\left( x_{1}+\eta
\right) \Delta t}+C_{2}\frac{\Omega _{b}}{2}e^{-i\left( x_{2}+\eta \right)
\Delta t}+C_{3}\frac{\Omega _{b}}{2}e^{-i\left( x_{3}+\eta \right) \Delta t}
\end{eqnarray*}
which are of the form (\ref{eqshift}). By similar calculations as above, one shows that a good approximation of the dynamics of the system is then obtained through replacing $\widetilde{\gamma}$ by its relevant part
\begin{equation}
\widetilde{\gamma}_{rel}=-\frac{\Omega_{a}}{2\Delta}\widetilde{\alpha}-\frac{\Omega_{b}}{2\Delta}
\widetilde{\beta} = e^{-i\eta\Delta t} \gamma_{rel}. \label{gamrelshift} 
\end{equation}
This expression differs from the rough adiabatic elimination result (\ref{gamshift}). As can be easily checked, however, when $\eta = O\left( \epsilon \right)$, (\ref{gamshift}) and (\ref{gamrelshift}) only differ by $O\left( \epsilon^{2} \right)$ terms which are not significant at the level of accuracy we consider. The rough elimination procedure thus appears as a practical (though physically not motivated) trick which works as the long as the origin of the energies lies in the ''neighbourhood'' of the midpoint energy $E_{0}$ of the two lower states, or, to be more explicit, as long as $E_{0}/\Delta = O\left( \epsilon \right)$.

The rigourous method we have employed here allowed us to clarify the applicability conditions of the rough procedure but it requires the exact solution of the Schr\"{o}dinger equation. In the next section, we present a more systematic and elegant approach based on the Green's function formalism which can be used to generalise the results presented here to more complicated level schemes \cite{BPM}.

\section{Rigorous adiabatic elimination in the Green's
function formalism \label{Green}\ \ }

\subsection{Overview of the Green's function formalism}

This overview summarizes the basic features of the Green's function and projection operator formalism first introduced in \cite{Feshbach} and extensively presented in
\cite{Cohen}. Given a system of Hamiltonian $H=H_{0}+V$, comprising a leading
part $H_{0}$ and a perturbation $V$, one defines the Green's function
$G\left(  z\right)  $ as follows\ \ \ \ \
\[
G\left(  z\right)  =\frac{1}{z-H}.
\]
The evolution operator of the system can then be derived through the formula%
\[
U\left(  t\right)  =\frac{1}{2\pi i}\int_{C_{+}\cup C_{-}}dze^{-izt/\hbar
}G\left(  z\right)
\]
where $C_{+}$ and $C_{-}$\ denote two parallel lines just above and below the
real axis, oriented from the right to the left and from the left to the right, respectively.\ \ \ \ \ 

Let $\mathcal{P}$ be the subspace spanned by some relevant eigenstates of $H_{0}$, and let $P$ and $Q=I-P$ be the orthogonal projectors on
$\mathcal{P}$ and $\mathcal{P}^{\perp}=\mathcal{Q}$, respectively. Then, one
can show \ \ \
\[
PG\left(  z\right)  P=\frac{P}{z-PH_{0}P-PR\left(  z\right)  P}%
\]
where the displacement operator $R\left(  z\right)  $ is defined by%
\[
R\left(  z\right)  =V+V\frac{Q}{z-QH_{0}Q-QVQ}V.
\]

\subsection{Application to the lambda system}

In our case, 
\[ 
P=\left\vert a\right\rangle \left\langle a\right\vert +\left\vert
b\right\rangle \left\langle b\right\vert, \quad Q=\left\vert e\right\rangle \left\langle e\right\vert, \quad
\widehat{H}_{0}=\hbar\left[
\begin{array}
[c]{ccc}%
-\frac{\delta}{2} & 0 & 0\\
0 & \frac{\delta}{2} & 0\\
0 & 0 & \Delta
\end{array}
\right], \quad \widehat{V}=\frac{\hbar}{2}\left[
\begin{array}
[c]{ccc}%
0 & 0 & \Omega_{a}^{\ast}\\
0 & 0 & \Omega_{b}^{\ast}\\
\Omega_{a} & \Omega_{b} & 0
\end{array}
\right]  
\]
whence 
\[
PG\left(  z\right)  P =\frac{P}{M\left(  z\right)  },\quad
M\left(  z\right)    \equiv z+\lambda\epsilon\hbar\Delta\sigma_{z}%
-\frac{\left(  \hbar\Delta\epsilon\right)  ^{2}}{z-\hbar\Delta}\left[
\begin{array}
[c]{cc}%
\left\vert \lambda_{a}\right\vert ^{2} & \lambda_{a}^{\ast}\lambda_{b}\\
\lambda_{a}\lambda_{b}^{\ast} & \left\vert \lambda_{b}\right\vert ^{2}%
\end{array}
\right]  .
\]
To compute 
\begin{equation}
PU\left(  t\right)P =\frac{1}{2\pi i}\int_{C_{+}\cup C_{-}}dze^{-izt/\hbar
}P/M\left(  z\right) \label{ProjEvol}
\end{equation}
one has to calculate the residues of $e^{-izt/\hbar}P/M\left(  z\right) $ at its poles which are the solutions of $\det\left[  M\left(  z\right)  \right]  =0$. Setting $x=\frac
{z}{\hbar\Delta}$, one gets%
\[
x^{3}-x^{2}-x\epsilon^{2}\left(  \lambda^{2}+\left\vert \lambda_{a}\right\vert
^{2}+\left\vert \lambda_{b}\right\vert ^{2}\right)  +\lambda\epsilon
^{2}\left(  \lambda+\left\vert \lambda_{a}\right\vert ^{2}\epsilon-\left\vert
\lambda_{b}\right\vert ^{2}\epsilon\right)  =0
\]
which coincides with (\ref{Eqfreq}) and thus leads to the same results $\left\{ x_{k} \right\}_{k=1,2,3}$ summarized in Table \ref{TabRes}. The associated residues $\left\{ \mathcal{R}_{k} \right\}_{k=1,2,3}$ are readily found to be
\begin{eqnarray*}
\mathcal{R}_{1} & = & e^{-i\Delta x_{1}t}\left[
\begin{array}
[c]{ll}%
1 & \frac{\lambda_{a}\lambda_{b}^{\ast}%
}{2\lambda}\epsilon \\
\frac{\lambda_{a}^{\ast}\lambda_{b}}{2\lambda}\epsilon  & 0
\end{array}
\right] +O\left(  \epsilon^{2}\right), 
\\ \mathcal{R}_{2}  & = & e^{-i\Delta x_{2}t}\left[
\begin{array}
[c]{ll}%
0  & -\frac{\lambda_{a}\lambda_{b}^{\ast}}%
{2\lambda}\epsilon \\
-\frac{\lambda_{a}^{\ast}\lambda_{b}}{2\lambda}\epsilon & 1
\end{array}
\right] +O\left(  \epsilon^{2}\right), \\
\mathcal{R}_{3} & = & O\left(  \epsilon^{2}\right)
\end{eqnarray*}
when $\lambda\neq0$, while when $\lambda=0$ 
\begin{eqnarray*}
\mathcal{R}_{1} & = & \frac{e^{-i\Delta x_{1}t}}{\left\vert \lambda
_{a}\right\vert ^{2}+\left\vert \lambda_{b}\right\vert ^{2}}\left[
\begin{array}
[c]{ll}%
\left\vert \lambda_{a}\right\vert ^{2} & \lambda_{a}\lambda_{b}^{\ast} \\
\lambda_{a}^{\ast}\lambda_{b}  &
\left\vert \lambda_{b}\right\vert ^{2}
\end{array}
\right]
+O\left(  \epsilon^{2}\right), 
\\ \mathcal{R}_{2} & = & \frac{e^{-i\Delta x_{2}t}}{\left\vert \lambda
_{a}\right\vert ^{2}+\left\vert \lambda_{b}\right\vert ^{2}}\left[
\begin{array}
[c]{ll}%
\left\vert \lambda_{b}\right\vert ^{2}  & -\lambda_{a}\lambda_{b}^{\ast} \\
-\lambda_{a}^{\ast}\lambda_{b} & \left\vert \lambda
_{a}\right\vert ^{2} 
\end{array}
\right] +O\left(  \epsilon^{2}\right),  \\
\mathcal{R}_{3} & = & O\left(  \epsilon^{2}\right) . 
\end{eqnarray*}

In both
cases, the third pole only contributes to the second order in $\epsilon$ to the evolution operator (\ref{ProjEvol}), whereas the first
two are $O\left(  1\right)$. To the first order, the last pole can thus
be omitted: this constitutes the so-called \textit{pole approximation}. \ 

A simple and straightforward way to implement the pole approximation is to replace $M\left(  z\right)$ by
\[
M^{\left(  0\right)  }\left(  z\right)  \equiv z+\lambda\epsilon\hbar
\Delta\sigma_{z}+\frac{\left(  \hbar\Delta\epsilon\right)  ^{2}}{\hbar\Delta
}\left[
\begin{array}
[c]{cc}%
\left\vert \lambda_{a}\right\vert ^{2} & \lambda_{a}^{\ast}\lambda_{b}\\
\lambda_{a}\lambda_{b}^{\ast} & \left\vert \lambda_{b}\right\vert ^{2}%
\end{array}
\right]
\]
in (\ref{ProjEvol}) which boils down to replacing $PR\left(  z\right)  P$ by $PR\left(
0\right)  P$ in the expression of $PG\left(  z\right)  P$. One readily shows that, as desired, the replacement of $PR\left(
z\right)  P$ by $PR\left(  0\right)  P$ in $PG\left(
z\right)  P$ discards the irrelevant pole and its associated residue, while
leaving the others unchanged (up to $O\left( \epsilon^{2} \right)$ terms).
Moreover $PG\left(  z\right)  P$ can
now be put under the form $PG\left(  z\right)P \simeq \frac{P}{z-H_{eff}}$
where $H_{eff}=PH_{0}P+PR\left(  0\right)P$ is a $2\times2$\ Hermitian
matrix, independent of $z$ which can be interpreted as the
effective Hamiltonian governing the dynamics of the reduced two-dimensional
system $\left\{  \left\vert a\right\rangle ,\left\vert b\right\rangle
\right\}  $: a simple calculation again leads to the expression (\ref{Hameff}) for $H_{eff}$.

Let us finish this section by some remarks. The calculations above have
been performed assuming that the energy $E_{0}$ at the midpoint of the two ground states spanning $\mathcal{P}$ is zero: if one shifts the
energies so that $E_{0}\neq0$, all the previous expressions and
calculations hold, up to the replacement of $z$ by its translated $z+E_{0}$,
as can be easily checked. The expression of the effective Hamiltonian thus
becomes\ \
\[
H_{eff}=PH_{0}P+PR\left(  E_{0}\right)  P,
\]
which is of course consistent with the particular case $\left(  E_{0}%
=0\right)  $ considered above.\ It is interesting to note that $H_{eff}$
always takes the same form, wherever one chooses the origin of the energies,
\textit{i.e.} the sum of the projected unperturbed Hamiltonian and the
operator $PRP$ evaluated at the middle energy of the subspace $\mathcal{P}$. Finally, let us also note that if $PRP$ is evaluated at an energy $E_{1}=E_{0}+O\left( \epsilon \right)$, the previous approximation remains valid, as the residues will only be affected by $O\left( \epsilon^{2} \right)$ terms, which are not significant at the level of accuracy considered. 

\section{Conclusion}

The goal of this article was to clarify the scheme usually employed to derive the effective two-dimensional Hamiltonian of a lambda system excited by off-resonant lasers: in particular we have reviewed two methods which enabled us to rigorously derive the effective dynamics of the system, up to terms to well-established order of magnitude in a small parameter; this study also allowed us to specify the applicability conditions of the rough elimination procedure. The second of these schemes relies on the Green's function formalism and the use of projectors on the different relevant subspaces of the state space of the system: this fairly general and elegant tool naturally leads to generalisations to more complicated multilevel systems, as shall be considered in a forthcoming paper.   

\begin{acknowledgments}
This work has been supported by ARO-DTO grant nr. 47949PHQC.

E.B. dedicates this work to the memory of Michel Barbara.
\end{acknowledgments}


\begin{thebibliography}{9}                                                                                                %
\bibitem {AE87}L.\ Allen and J.H. Eberly, "Optical Resonance and Two-Level
Atoms", Dover Publications, Inc., New York (1987).

\bibitem {DC89} J. Dalibard, C. Cohen-Tannoudji,, J.O.S.A. B \textbf{6}, 2023 (1989) 

\bibitem {Wineland}B. E. King, C. S. Wood, C. J. Myatt, Q. A. Turchette, D. Leibfried, W. M. Itano, C. Monroe, and D. J. Wineland, Phys. Rev. Lett. \textbf{81}, 1525 (1998).

\bibitem {Saffman}M. Saffman and T. G. Walker, Phys. Rev. A \textbf{72},
022347 (2005).

\bibitem {Ima}A. Imamoglu, D. D. Awschalom, G. Burkard, D. P. DiVincenzo, D. Loss, M. Sherwin, and A. Small, Phys. Rev. Lett. \textbf{83}, 4204 (1999).


\bibitem{Feshbach}H. Feshbach, Ann. Phys. (N.Y.) \textbf{5}, 357 (1958); H. Feshbach, Ann. Phys. (N.Y.) \textbf{19}, 287 (1962);  

\bibitem {Cohen}C. Cohen-Tannoudji, J. Dupont-Roc, G. Grynberg, Atom-Photon Interactions: Basic Processes and applications (Wiley, New-York, 1992).

\bibitem {BPM}E. Brion, L.H. Pedersen, and K. M\o lmer, Adiabatic Elimination in Multilevel Systems, submitted.
\end{thebibliography}
\end{document}